\documentclass[12pt]{revtex4}

\usepackage{amsmath,amssymb,graphicx}
\usepackage{epsfig}
\usepackage{textcomp}
\usepackage{color}
\usepackage{cases}
\usepackage{epstopdf}
\usepackage{multirow}
\usepackage{hyperref}
\hypersetup{pdfpagemode=UseNone,colorlinks=true,linktoc=page,pdfborder={0 0 0},linkcolor=blue,citecolor=blue}
\usepackage{amstext}
\usepackage{bm}
\usepackage{color}

\begin{document}
\title{Time Distribution for Persistent Viral Infection}
\author{Carmel Sagi and Michael Assaf}

\affiliation{Racah Institute of Physics, Hebrew University of Jerusalem, Jerusalem 91904, Israel}
\begin{abstract}
We study the early stages of viral infection, and the distribution of times to obtain a persistent infection. The virus population proliferates by entering and reproducing inside a target cell until a sufficient number of new virus particles are released via a burst, with a given burst size distribution, which results in the death of the infected cell. Starting with a 2D model describing the joint dynamics of the virus and infected cell populations, we analyze the corresponding master equation using the probability generating function formalism. Exploiting time-scale separation between the virus and infected cell dynamics, the 2D model can be cast into an effective 1D model. To this end, we solve the 1D model analytically for a particular choice of burst size distribution. In the general case, we solve the model numerically by performing extensive Monte-Carlo simulations, and demonstrate the equivalence between the 2D and 1D models  by measuring the Kullback-Leibler divergence between the corresponding distributions. Importantly, we find that the distribution of infection times is highly skewed with a ``fat" exponential right tail. This indicates that there is non-negligible portion of individuals with an infection time, significantly longer than the mean, which may have implications on when HIV tests should be performed.
\end{abstract}

\maketitle

\section{Introduction}\label{introduction}
Dynamics of infectious diseases, and in particular, the early stages of infection, have attracted much attention in recent years, see \textit{e.g.}, Refs.~\cite{anderson1992infectious,nowak2000virus,helbing2015saving,wang2016statistical}. An important subcategory of such diseases are those that are transmitted via a viral infection. This work focuses on the investigation of a prototypical stochastic model, which describes the early infection stages of various viral infections such as the Human immunodeficiency virus type 1 (HIV-1 or simply HIV), or the hepatitis C virus (HCV)~\cite{perelson1996hiv,pearson2011stochastic}.

The HIV attacks CD4$^+$ cells which belong to the immune system. An HIV virion (virus particle, the infective extracellular form of the virus)  attaches to a healthy CD4$^+$ cell, and enters it. Inside the cell the virion uses reverse transcriptase and changes its RNA to DNA, allowing it to penetrate the cell's nucleus. Once integrated, the virion uses the cell's nucleus to produce new virions via bursts (or in a continuous manner)~\cite{pearson2011stochastic}. This process eventually kills the infected cell and releases the virions. In the first stages of infection, before the immune system responds, the dynamics of this process is highly noisy. On the one hand, the virions have a nonzero degradation rate which may lead to extinction of the virus, even without an intervention of the immune system. The exact reasons for virion degradation are unknown; the conjecture is that it mainly results from binding to cells or non specific immune elimination \cite{perelson1996hiv,li2009visualizing}.
On the other hand,  given that the virions have not gone extinct, the noisy dynamics yields a high variability in the infection time -- the time it takes the virions to infect a sufficient number of cells such that viral extinction is highly unlikely. Thus, in many cases the \textit{mean} time to infection, which can be calculated using standard techniques, becomes inadequate for reliably predicting the fraction of infected individuals within a population up to a given time.

The dynamics of infected cells and virions can be described by the following reaction set
\begin{equation}
V\overset{\beta}{\rightarrow}I,\quad I\overset{\alpha}{\rightarrow}kV,\quad
V\overset{1}{\rightarrow}\emptyset.\label{1}
\end{equation}
Here $I$ represents the number of infected cells, $V$ represents the number of virions and $\emptyset$ denotes an empty set. The first reaction describes penetration of a virion into a cell and its infection, which occurs at a rate $\beta$. The second reaction, which occurs at a rate of $\alpha$, describes a burst of virions which releases $k$ new virions (see below), and kills the host cell. The third reaction describes degradation of virions at a rate of 1 per individual. Note, that here we rescale all rates by the virion degradation rate, which is ${\cal O}(10)$ per day~\cite{ramratnam1999rapid}, and we neglect spontaneous degradation of infected cells, whose rate is negligible~\cite{selliah2003t}.

Model~(\ref{1}) describes the so-called "exposure phase", which corresponds to the  linear regime -- the early stages of infection. That is, we assume that (i) the infected cell number is undetectable to the immune system, and (ii) the population of uninfected cells, which are susceptible to virion infection, is constant~\cite{footnote1}. Furthermore, the model is generic and can describe any viral infection that behaves similarly as the HIV, such as the HCV \cite{ramratnam1999rapid,martinez2015similarities} or even the influenza virus \cite{baccam2006kinetics,footnote2}. Hence, the model can predict the distribution of infection times (DIT) for  a wide variety of viruses that proliferate in this manner.

The stochastic process, described by Eq.~(\ref{1}), has two possible outcomes: extinction of the virions and infected cells, or unlimited growth (or proliferation) of the number of virions and infected cells which is ultimately arrested by non-linear terms. In the latter case, the infection is called persistent if the number of infected cells reaches a given threshold $N_0\gg 1$, from which the extinction probability is vanishingly small. In this work we omit the non-linear terms~\cite{footnote3} focusing on the linear regime, and demonstrate that the stochastic nature of the viral dynamics plays a crucial role in determining the statistics of infection times.

The number of released virions per cell depends on many factors including the different phases of infection \cite{rusert2004quantification}.  Chen et al. have shown that in simian immunodeficiency virus (SIV) infection -- a monkey-infecting virus similiar to HIV -- about 50,000 virions are released upon a death of a cell \cite{chen2007determination}. However, a large fraction of these virions are not sufficiently mature and thus are non infectious, while others die before they infect new healthy cells \cite{briggs2009structure}. As a result, the fraction of released \textit{infectious} virions during the first stages of infection are estimated to be 1 in $10^3$ to $10^4$ virions  \cite{bourinbaiar1994ratio,marozsan2004relationships,dimitrov1993quantitation}.  In our model, since we are interested in the dynamics of infected cells, we only account for the infectious virions and thus, we consider $k$ values which are on the order of $1-100$.

In recent years, the early stages of HIV dynamics have been extensively investigated using different methods. Tan and Wu \cite{tan1998stochastic} derived a stochastic infection model and used Monte-Carlo simulations to study time-dependent distributions of the numbers of virions and infected cells. They have also shown that viral extinction probability is finite, see also Refs.~\cite{pearson2011stochastic,kamina2001stochastic,heffernan2005monte}.  Pearson et al.~\cite{pearson2011stochastic} also computed the DIT numerically. Tuckwell and Le Corfec~\cite{tuckwell1998stochastic} introduced a modified infection model by incorporating white noise into the deterministic model. They have shown that the virion growth does not depend on the initial viral population, while Lee et al.~\cite{lee2009modeling} determined the effect of the initial conditions on the probability to reach a viable infection. Chaudhury et al.~\cite{chaudhury2012spontaneous} investigated the role of demographic noise in early stages of viral infection by numerically solving the master equation. Furthermore, Noecker et al.~\cite{noecker2015simple} studied early phases of infection, by accounting for an additional phase in which infected cells do not release virions; they also computed the DIT numerically. However, despite the large body of work in this field, to the best of our knowledge, the DIT to reach a persistent infection has not been studied analytically so far.

In this work we study the DIT to reach persistent infection in a 2D model, which accounts for the interplay between the dynamics of infected cells and virions. To this end we use extensive Monte-Carlo simulations as well as analytical methods such as the probability generating function formalism. In the limit where the viral dynamics is fast compared to that of the infected cells, we derive an effective 1D model which includes the dynamics of the infected cells only. We demonstrate, both analytically and numerically, that the 1D model can reproduce the DIT of the 2D model with a high accuracy, for a generic bursty scheme.

The paper is organized as follows. In Sec.~\ref{Theoretical Analysis} we present the theoretical analysis of the model. In Sec.~\ref{Determinstic Approach} we present the deterministic approach. The stochastic approach is described in Sec.~\ref{Stochastic Approach: 2D model} by using the probability generation function formalism. In Sec.~\ref{Subsection 1D model} we find an effective 1D model, while in Sec.~\ref{BLimit} the bifurcation limit is considered. Section~\ref{Numerical Analysis} is dedicated to presenting the numerical Monte-Carlo simulation results. In Sec.~\ref{Comparison 1D 2D Models} we show comparisons between the 1D and 2D models, while in Sec.~\ref{Statistics Infection} we find the DIT semi-empirically. Finally, in Sec.~\ref{Implications Realistic Population}  we discuss the implications of our model on realistic populations and employ our model to HCV. We conclude our study in Sec.~\ref{Conclusion Discussion}.

 \section{Theoretical Analysis}\label{Theoretical Analysis}
\subsection{Deterministic approach}\label{Determinstic Approach}
Our starting point is model (\ref{1}) describing the dynamics of virions ($V$) and infected cells ($I$). When the populations of infected cells and virions are large, one can neglect demographic noise and write down  the dynamics in terms of a set of coupled ordinary differential equations. These describe the time evolution of the \textit{mean} number of virions and infected cells. Using model~(\ref{1}) we arrive at
\begin{equation}
\dot{I}=-\alpha I + \beta V,\quad\dot{V}= -\left(\beta+1\right) V  + \alpha k I,\label{2}
\end{equation}
where $I$ represents the mean number of infected cells, $V$ represents the mean number of virions, and $k$ represents the number of released infectious virions. As a reminder, these equations account for three reactions: infection of a cell at a rate of $\beta$, death of an infected cell and release of $k$ virions at rate $\alpha$, and virion degradation at a rate of 1 per individual.

In realistic systems, $\alpha$ is small compared to the other rates, as the infected cells die at a relatively slow rate compared to other processes, see e.g. Refs. \cite{pearson2011stochastic,nowak2000virus}. As a result, we assume henceforth that $\alpha\ll 1$, while $\beta={\cal O}(1)$. In the literature, one is interested in the infection dynamics starting from a single virion  at $t=0$ \cite{pearson2011stochastic,tuckwell1998stochastic}. %abrahams2009quantitating,haaland2009inflammatory,kearney2009human,keele2009low,fischer2010transmission}.
Using the initial condition, $V(t=0)=1$ and $I(t=0)=0$, and for $\alpha\ll 1$, the solution to~(\ref{2}) reads
\begin{equation}
I\simeq \frac{\beta}{\beta+1}\left[e^{\lambda_{1} t}-e^{\lambda_{2} t}\right],\;\;\;\;\;\;V\simeq\frac{\beta k \alpha}{\left(\beta +1\right) ^2}e^{\lambda_{1} t}+\left[1-\frac{\beta k \alpha}{\left(\beta +1\right) ^2}\right]e^{\lambda_{2} t},
\end{equation}
where $\lambda_{1}$, $\lambda_{2}$ are the eigenvalues, satisfying
\begin{equation}
\lambda_{1}\simeq  \alpha\left(\frac{\beta k }{\beta +1}-1\right),\quad\lambda_{2}\simeq-\left(1+\beta+\frac{\beta k \alpha}{\beta +1 }\right).
\end{equation}
Let us define the basic reproductive ratio, $R\equiv\beta k/\left(\beta+1\right)$~\cite{pearson2011stochastic}. In epidemiology, $R$ is used to study the rate of spread of an infectious disease~\cite{anderson1992infectious}, while here $R$ represents the number of new cells that a single infected cell will infect during its lifespan. One can show (also in the epidemiological context) that $R>1$ describes a persistent infection, while for $R\le1$ the infection dies out deterministically. Indeed in the latter case both eigenvalues are negative, while for $R>1$, $\lambda_{1}>0$, and the populations of $I$ and $V$ grow exponentially. Furthermore, since $\alpha\ll1$, one has $\lambda_{1}\ll\left|\lambda_{2}\right|$, and thus, at long times $t\gtrsim \mathcal{O}\left(1/\lambda_{1}\right)$ the dynamics is solely governed by $\lambda_1$. At this point it is convenient to define $\lambda\equiv\lambda_{1}$ as the (small) positive eigenvalue. By further rescaling time $\tilde{t}=\lambda t$, the solution to Eq. (\ref{2}) at long times satisfies
\begin{equation}
I=\frac{\beta}{\beta+1}e^{\tilde{t}},\quad V=\frac{\beta k\alpha}{\left(\beta+1\right)^2}e^{\tilde{t}}.\label{3}
\end{equation}
Henceforth, we omit the $\sim$ above $t$ and measure time in units of the inverse of $\lambda$ times the virus degradation rate.

An important feature of Eqs.~(\ref{2}) is that for $\lambda\sim\alpha\ll1$, a timescale separation occurs between the rapidly-varying $V$ and the slowly-varying $I$ \cite{haken1983synergetics,constable2014fast}. Indeed, defining $\epsilon=\lambda/\left(\beta+1\right)\ll1$, and using the normalized time, rate equations~(\ref{2}) become
\begin{equation}
\frac{\mathrm{d}I}{\mathrm{d}t}=-\frac{\alpha}{\lambda}I+\frac{\beta}{\lambda}V,\quad
\epsilon\frac{\mathrm{d}V}{\mathrm{d}t}=-V+\frac{\alpha k}{\beta+1}I.\label{4}
\end{equation}
This indicates that in the limit of $\epsilon\rightarrow0$, $\epsilon dV/dt$ vanishes, and $V$ becomes enslaved to $I$. As a result, $V$ is expected to rapidly fluctuate around a slowly varying trend, $V(I)\simeq\alpha k I/\left(\beta+1\right)$, see Fig.~\ref{fig:1} and below.

\subsection{Stochastic approach: 2D model}\label{Stochastic Approach: 2D model}
The deterministic approach predicts an exponential growth of $I$ and $V$ for $R>1$, but it neglects demographic noise due to the discreteness of individuals and stochastic nature of the reactions. This noise brings about qualitative changes in the dynamics. For example it allows for the virion and infected cell populations to go extinct. On the other hand, if extinction does not occur, the time to reach $N_0$ infected cells (where $N_0$ defines a persistent infection, see Sec.~\ref{introduction}) is strongly varying across the different stochastic realizations, which results in a non-trivial DIT, see below. Figure~\ref{fig:1} shows an example of three stochastic realizations of process~(\ref{1}) and the deterministic solution [Eq.~(\ref{3})]. One can see the significant variability in the infection times across these realizations.
\begin{figure}[ht]
	\centering
	\includegraphics[width=.7\linewidth]{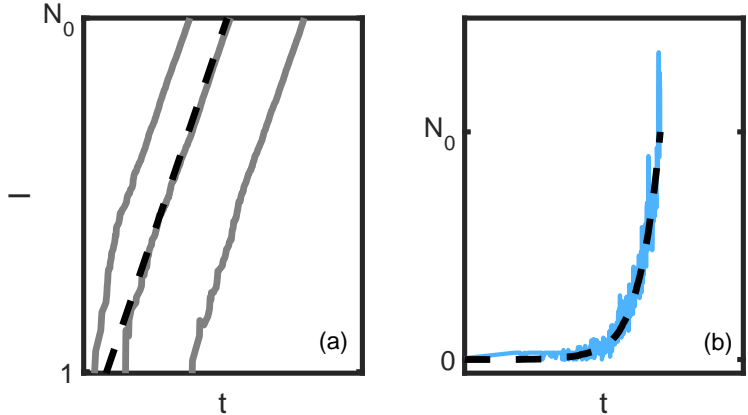}
	\caption{(a) Heuristic plot of $I(t)$ (on a semi-log scale) for three realizations of process~(\ref{1}): The left, middle and right solid lines represent a fast, an average, and a slow realization, respectively. The dashed line denotes the deterministic solution given by Eq. (\ref{3}). In (b) shown is $I(t)$ along the average realization (dashed line), and $V(t)$ multiplied by  $\left(\beta+1\right)/\left(\alpha k\right)$ (solid line), demonstrating that $V$ is an enslaved variable to $I$. In both panels $\beta=2$, $k=5$, $\alpha=0.01$, $N_0=10^4$.}
	\label{fig:1}
\end{figure}

Apart from demographic noise, our model includes an \textit{additional} source of uncertainty -- the number of virions produced per burst.
To account for the stochastic dynamics, we write down the master equation which describes the evolution of the probability $P_{n,m}\left(t\right)$ to find $n$ infected cells and $m$ virions at time $t$ \cite{be2016effect,be2016rare,be2017enhancing}
\begin{eqnarray}
\dot{P}_{n,m}\left(t\right)\!&=&\!\frac{\alpha}{\lambda}\!\left[\!\left(n\!+\!1\right)\!\sum_{k\!=0}^{\infty}D\left(k\right)P_{n+1,m-k}\left(t\right)\!-nP_{n,m}\left(t\right)\!\right]\nonumber\\
&+&\frac{\beta}{\lambda}\left[\left(m+1\right)P_{n-1,m+1}\left(t\right)-mP_{n,m}\left(t\right)\right]+\frac{1}{\lambda}\left[\left(m+1\right)P_{n,m+1}\left(t\right)-mP_{n,m}\left(t\right)\right].\label{5}
\end{eqnarray}
Here $D\left(k\right)$ denotes the burst size distribution (BSD)~\cite{be2016effect,be2016rare,be2017enhancing} -- the probability distribution for $k$ virions to be released upon the death of the cell. Multiplying Eq.~(\ref{5}) once by $n$ and summing over all $n$'s, and once by $m$ and summing over all $m$'s, and recalling that $\langle n\rangle=\sum_{n,m=0}^{\infty}nP_{n,m}\left(t\right)$, and $\langle m\rangle=\sum_{n,m=0}^{\infty}mP_{n,m}\left(t\right)$, we arrive at the deterministic rate equations [Eqs.~(\ref{2})] for the mean number of infected cells and virions.

Throughout this work we will focus on two types of BSDs: the $K$-step distribution, $D\left(k\right)=\delta_{k, K}$, and the geometric distribution, $D\left(k\right)=q^k\left(1-q\right)$, where $k=0,1,2,\dots$, $q=\langle k\rangle/\left(\langle k\rangle+1\right)$ is the probability to release a single virion, and $\langle k\rangle$ is the average number of virions released. In the latter case, for which the number of released virions is a-priori unknown, the effective reproductive ratio equals  $R=\beta\langle k\rangle/\left(\beta+1\right)$. A special case for the $K$-step distribution is the so-called single-step reaction (SSR) for $K=2$, which will be analyzed in detail. Since the typical number of released infectious virions is at most ${\cal O}(100)$, we have taken  $K$ and $\langle k\rangle$ to be up to 100.

Before computing $P_{n,m}\left(t\right)$ and the DIT, let us first study the extinction scenario, by computing the quantity $\Pi_{n,m}$ -- the probability that starting with $n$ infected cells and $m$ virions, both populations will eventually undergo extinction. It can be shown that $\Pi_{n,m}$ satisfies the following recursive equation~\cite{gardiner2004handbook,feller1968introduction,redner2001guide,parzen1960modern,mosteller1965fifty,pearson2011stochastic}
\begin{equation}
\Pi_{n,m}=\frac{\alpha n}{ z}\sum_{k=0}^{\infty}D\left(k\right)\Pi_{n-1,m+k}+\frac{m\beta}{z}\Pi_{n+1,m-1}
+\frac{m}{z}\Pi_{n,m-1},\label{6}
\end{equation}
where $z=\beta m+\alpha n+m$ is the sum of rates to leave a state with $n$ infected cells and $m$ virions. Here, starting from $n$ infected cells and $m$ virions, the probability of extinction equals the sum of three terms: (i) probability of extinction starting from $n-1$ infected cells and $m+k$ virions times the probability that an infected cell has died and released $k$ new virions; (ii) probability of extinction starting from $n+1$ infected cells and $m-1$ virions times the probability that a virion has been absorbed by a cell resulting in its infection, and (iii) probability of extinction starting from $n$ infected cells and $m-1$ virions times the probability that a virion died.

Equation (\ref{6}) is a 2D recursive equation, and solving it is equivalent to solving a partial differential equation. Yet, the fact that the transition rates are linear in $n$ and $m$ allows one to find a solution in a straightforward manner, by noting that the probability of extinction starting from $n$ infected cells is equivalent to the probability of extinction of $n$ separate realizations, each starting with one infected cell. The same goes for the virion population. As a result, we can look for the solution as $\Pi_{n,m}=\rho_I^n\rho_V^m$, where $\rho_I$ and $\rho_V$ are the probabilities that an infection initiated with a single infected cell and a single virion, respectively, will result in extinction. The solution has to satisfy the boundary conditions  $\Pi_{\infty,\infty}=0$, since $\rho_I,\rho_V<1$, and $\Pi_{0,0}=1$, as for $m=n=0$ extinction has already occurred. Substituting this solution into Eq.~(\ref{6}), one obtains a characteristic polynomial equation for $\rho_I$ and $\rho_V$,
%\begin{equation}
%\rho_I^n\rho_V^m\!=\!\frac{\alpha n }{z}\sum_{k=0}^{\infty}D\!\left(k\right)\rho_I^{n-1}\rho_V^{m+k}\!+\!\frac{m\beta}{z}\rho_I^{n+1}\!\rho_V^{m\!-\!1}\!+\!\frac{m}{z}\rho_I^n\rho_V^{m\!-\!1},\
%\end{equation}
whose solution, for arbitrary $D\left(k\right)$, yields \cite{pearson2011stochastic}
\begin{eqnarray}
\rho_I=\sum_{k=0}^{\infty}D\left(k\right)\rho_V^k,\;\;\;\;\;\;\;\;\sum_{k=0}^{\infty}\frac{\beta}{\beta+1}D\left(k\right)\rho_V^k-\rho_V+\frac{1}{\beta+1}=0.\label{7}
\end{eqnarray}
For the $K$-step BSD, $\rho_I$ and $\rho_V$ satisfy \cite{pearson2011stochastic}:
\begin{eqnarray}
\rho_I=\rho_V^{K},\quad0=\rho_V^{K}-\frac{\beta+1}{\beta}\rho_V+\frac{1}{\beta},\label{8}
\end{eqnarray}
where these equations can be solved numerically for any $K$. Analytical solutions are available for $K\le4$; For example, for $K=2$, $\rho_V=1/\beta$, while for $K=3$, $\rho_V=\sqrt\beta\sqrt{\beta+4}/\left(2\beta\right)-1/2$. For the geometric BSD we find
\begin{equation}
\rho_I=\frac{1+\beta}{\beta \langle k\rangle },\;\;\;\;\rho_V=\frac{1+\beta + \langle k\rangle }{\langle k\rangle\left(1+\beta\right)}.
\end{equation}

Note that for $R<1$, one can show that $\Pi_{n,m}=1$, namely, virion extinction is guaranteed \cite{pearson2011stochastic}.
\begin{figure}[ht]
	\centering
	\includegraphics[width=.7\linewidth]{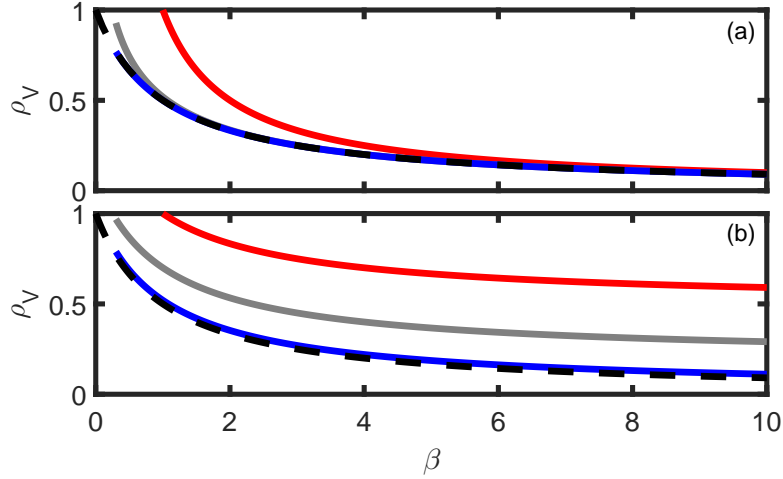}
	\caption{The extinction probability starting with a single virion (and zero infected cells), $\rho_V$, as a function of the cell infection rate $\beta$. (a) The case of $K$-step BSD with $K$=2, 5, 50 (top to bottom). (b) The case of geometric BSD for $\langle k\rangle=2,5,50$ (top to bottom). In both panels, for large $K$, one can see that $\rho_V$ approaches $1/\left(1+\beta\right)$ (denoted by the dashed lines). One can also see that for a given $\langle k\rangle=K$, $\rho_V$ for the geometric BSD is higher than that of the $K$-step BSD, as in the former case there is a nonzero probability to release zero virions in a burst event, see text.}
	\label{fig:2}
\end{figure}
In Fig.~\ref{fig:2} we present $\rho_V$ for both the $K$-step and geometric BSDs as a function of $\beta$ for different values of $K$ and $\langle k\rangle$, respectively. In both cases, $\rho_V$ decreases  (and the infection probability increases) with increasing $K$ and $\langle k\rangle$, as more virions are released per burst, and more cells are infected due to the higher number of virions. $\rho_V$ also decreases as $\beta$ increases.  For the $K$-step BSD, in the limit of large $\beta$, one can show that the solution for $\rho_V$ approaches $1/\beta$ for any $K\ge 2$. Alternatively, for any $\beta$, when $K$ is large, the second of Eqs. (\ref{8}) can be reduced to $\rho_V=1/\left(\beta+1\right)$ as $\rho^K$ is negligible, see Fig.~(\ref{fig:2}). Indeed, if infection starts with a single virion there are two possibilities: the virion can infect a cell with probability $\beta/\left(\beta+1\right)$ or  die with probability $1/\left(\beta+1\right)$. Thus, for $K\gg 1$, if extinction does not occur initially, persistent infection will almost certainly occur.  Moreover, for any $\langle k\rangle = K$, the extinction probability for the geometric BSD is higher than that of the $K$-step BSD, as there is a possibility for zero virions to be released in the former case. Note  that when $\langle k\rangle\gg \beta$ the geometric BSD solution also approaches $1/\left(\beta+1\right)$.

\subsubsection{Probability generating function formalism}\label{PG function formalism}
In this subsection we use the probability generating function (PGF) formalism~\cite{gardiner2004handbook} to analyze master equation (\ref{5}) and to find the DIT. We do so by calculating the probability distribution of finding $n$ infected cells at time $t$ regardless of the number of virions, which is then used to obtain the DIT to reach $N_0$ cells. Previous works \cite{pearson2011stochastic,noecker2015simple} have shown  numerical calculations for the DIT, and here we find the distribution analytically.

The PGF for a two-variable probability distribution function has the form
\begin{equation}
G\left(x,y,t\right)=\sum_{n,m=0}^{\infty}x^ny^mP_{n,m}\left(t\right),
\end{equation}
where $x$ and $y$ are auxiliary variables representing the infected cells and virions respectively. Importantly, the PGF encodes the probability distribution, which can be found by Taylor-expanding $G$ about $x=y=0$~\cite{gardiner2004handbook}:
\begin{equation}
P_{n,m}\left(t\right)=\frac{1}{n!m!}\left.\frac{\partial^{n+m} G\left(x,y,t\right)}{\partial x^n\partial y^m}\right|_{x=0, y=0}.\label{9}
\end{equation}

To find the PGF, we multiply master equation (\ref{5}) by $x^ny^m$ and sum over all $n'$s and $m'$s, arriving at a first-order partial differential equation
\begin{equation}
\frac{\partial G}{\partial t}=\frac{\partial G}{\partial y}\frac{\left(\beta x-\beta y+1-y \right)}{\lambda}\\
+\frac{\partial G}{\partial x}\frac{\alpha}{\lambda}\left[\sum_{k=0}^{\infty}y^kD\left(k\right)- x\right]\!.\label{10}
\end{equation}
This equation cannot be solved analytically due to the coupling between $x$ and $y$. To proceed, we exploit the smallness of the parameter $\alpha$ which gives rise to  time-scale separation between $V$ and $I$. Indeed, we notice that the coefficient multiplying $\partial G/\partial y$ diverges when $\lambda\sim\alpha\rightarrow0$, while the other coefficients are $\mathcal{O}\left(1\right)$. As a result, to regularize Eq.~(\ref{10}) we demand that the numerator in the coefficient of $\partial G/\partial y$ vanish. Thus, we demand that $\beta x-\beta y+1-y=0$, which yields $y=\left(1+\beta x\right)/\left(1+\beta\right)$. This is the analogous relation in the stochastic picture between $V$ and $I$, to the deterministic relation between these variables, see below Eq.~(\ref{4})~\cite{footnote4}. Using this relation between $y$ and $x$,  Eq. (\ref{10}) becomes
\begin{equation}
\frac{\partial G}{\partial t}=\frac{\partial G}{\partial x}\frac{\alpha}{\lambda}\left[\sum_{k=0}^{\infty}\left(\frac{1+\beta x }{1+\beta  }\right)^kD\left(k\right)- x\right].\label{11}
\end{equation}
Note, that $G$ is now only a function of $x$. In fact, by eliminating the $y$ variable, finding $G(x,t)$ yields the probability to find $n$ infected cells at time $t$ regardless of the number of virions, which is exactly what we are after.

\subsubsection{The case of single-step reaction}\label{The Case SSR}
\begin{figure}[ht]
	\centering
	\includegraphics[width=.6\linewidth]{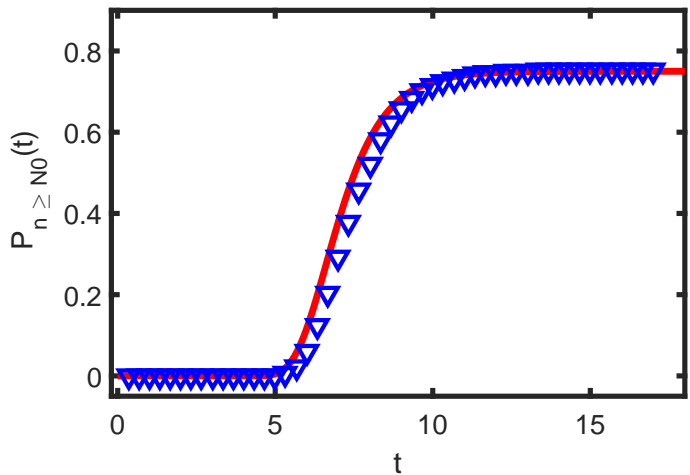}
	\caption{Shown is the cumulative distribution -- the probability to reach at least $N_0$ infected cells -- as a function of the rescaled time, in the case of the SSR. Here, the analytical result given by Eq.~(\ref{cumdist}) (solid line) is compared with results of numerical simulations (triangles). At long times, $t\gg 1$, the cumulative distribution approaches the infection probability, $Q_{2D}$. Parameter values are $N_0=10^3$, $\alpha=0.1$ and $\beta=2$.  }
	\label{fig:3}
\end{figure}

We now solve Eq.~(\ref{11}) for the SSR case, for which $D(k)=\delta_{k,2}$. Here, Eq.~(\ref{11}) becomes
\begin{equation}
\frac{\partial G}{\partial t}=\frac{\partial G}{\partial x}\frac{\alpha}{\lambda}{\left[\left(\frac{1+\beta x }{1+\beta}\right)^2-x\right]}.\label{12}
\end{equation}
This equation has to be solved with the initial condition $P_n\left(t=0\right)=\delta_{n,1}$, namely, starting from a single infected cell. Using the definition of $G$ in 1D, $G\left(x,t\right)=\sum_{n=0}^{\infty}x^n P_n\left(t\right)$, the initial condition becomes $G\left(x,t=0\right)=x$. In addition, conservation of probability yields the boundary condition $G\left(x=1,t\right)=\sum_{n=0}^{\infty}P_n\left(t\right)=1$.
With these initial and boundary conditions, employing the method of characteristics~\cite{gardiner2004handbook}, the solution of Eq.~(\ref{12}) reads
\begin{equation}
G\left(x,t\right)=\frac{1-Q_{2D}-e^t (x-1)\left(Q_{2D}-1\right)-x}{1-Q_{2D}+e^t (x-1)-x}.\label{13}
\end{equation}
Here we have defined the survival probability starting from one cell, $Q_{2D}=1-\rho_I=1-1/\beta^2$, where the subscript 2D denotes the two dimensional model~(\ref{1}). To compute $P_n\left(t\right)$ -- the probability to find $n$ infected cells at time $t$ -- we use the 1D version of~(\ref{9}), which yields
\begin{equation}
P_n\left(t\right)=\frac{Q_{2D}^2e^t}{\left(e^t-1+Q_{2D}\right)^2}\,\left(\frac{e^t-1}{e^t-1+Q_{2D}}\right)^{n-1}.
\end{equation}
Being interested in the long time behavior such that $t\gg1$, we finally arrive at
\begin{equation}
P_n\left(t\right)\simeq Q_{2D}^2 e^{-n Q_{2D} e^{-t}
   -t}.\label{PNT}
\end{equation}
We now use this quantity to compute the DIT to reach a given $N_0$, $P_t\left(N_0\right)$. Using the fact that the probability contained in a differential area is invariant under change of variables, we have $\left|P_n\left(t\right)\mathrm{d}n\right|=\left|P_t\left(n\right)\mathrm{d}t\right|$, or $P_n(t)|dn/dt|=P_t(n)$, where at long times, we can estimate $dn/dt\simeq n$, since $n(t\gg 1)\simeq e^{t}$. Normalizing the resulting distribution such that $\int_0^{\infty} P_t(N_0)dt=1$, we find the DIT to be
\begin{equation}
P_t\left(N_0\right)\simeq\frac{Q_{2D} \left(N_0-1\right) }{1-e^{Q_{2D} (1-N_0)}}e^{-Q_{2D}N_0e^{-t}-t}.\label{Time Distribution 2D}
\end{equation}
 This expression is one of our main results. It is also interesting to compute the probability to have at least $N_0$ infected cells at time $t$, i.e., the cumulative distribution of Eq. (\ref{PNT}). Summing $P_n\left(t\right)$ over $n$ from $N_0$ to $\infty$ we find:
\begin{equation}\label{cumdist}
P_{n\geq N_0}\left(t\right)=Q_{2D} e^{-Q_{2D} N_0
   e^{-t}}.
\end{equation}
At $t\rightarrow\infty$, its value approaches the infection probability, $Q_{2D}$. The analytical expression for the cumulative distribution excellently agrees with numerical simulations, see Fig.~\ref{fig:3}.

\begin{figure}[ht]
	\centering
 	\includegraphics[width=.65\linewidth]{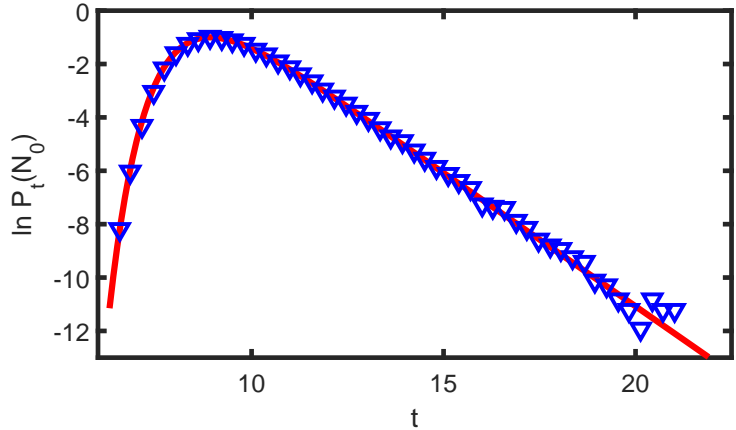}
	\caption{Shown is the distribution of infection times to reach $N_0$ infected cells, as a function of the rescaled time, in the case of the SSR. Analytical results [Eq.~(\ref{Time Distribution 2D})] (solid line) are compared with numerical simulations (triangles), for $\beta=2$, $\alpha=0.01$ and $N_0=10^4$.}
	\label{fig:4}
\end{figure}

Our numerical simulations of model (\ref{1}) were carried out by using the Gillespie algorithm \cite{gillespie1977exact}. A comparison between the analytical and numerical DITs for the case of SSR is shown in Fig. \ref{fig:4}, and excellent agreement is observed. In order to obtain the numerical DIT, we have determined an infection threshold and binned the infection times from all realizations that reached infection, using bins of size $\Delta t$. The corresponding error in each bin is approximately given by $1/\sqrt{MQ_{2D}P_t(N_0)\Delta t}$, where $M$ is the total number of realizations such that $MQ_{2D}$ is the number of realizations that reached infection, and $P_t(N_0)\Delta t$ is the probability to get infected between time $t$ and $t+\Delta t$. We made sure that in all our plots this error was at most 10$\%$, and the size of the symbols in all figures accounts for this error.

\subsection{Effective 1D model}\label{Subsection 1D model}
So far, we have seen that for $\alpha\ll 1$, the dynamics of the virions is enslaved to that of the infected cells. We now exploit this property, which is generic and holds for any BSD, to reduce our 2D model into an effective 1D model, which can be analyzed, both analytically and numerically, in a simpler manner. Indeed, in the limit of $\alpha\rightarrow0$, the dynamics of the virions is instantaneous compared to that of the infected cells. As a result, one can write $I\rightarrow kV\rightarrow kI$. Assuming that the cells also degrade at some rate due to the degradation of virions, to be determined below, the effective 1D dynamics of the infected cells can be described by the following two reactions
\begin{equation}
I\overset{\gamma}{\rightarrow}kI,\quad I\overset{\delta}{\rightarrow}\emptyset.\label{1Dmodel}
\end{equation}
Here $\gamma$ and $\delta$ are the effective burst and degradation rates, respectively, yet to be determined, while $k$ is the number of infected cells created per burst event, drawn from the same BSD, $D\left(k\right)$, as in the 2D model. To find $\gamma$ and $\delta$, we demand that the growth rate of the infected cell population, and its survival probability, coincide between the 1D and 2D models.

Equation~(\ref{1Dmodel}) yields the following rate equation: $\dot{I}=\left[\left(k-1\right)\gamma-\delta\right]I$, whose solution is $I=e^{\left(\left[k-1\right]\gamma-\delta\right)t}$. Thus, to match between the 1D and 2D growth rates, we demand that $\left(k-1\right)\gamma-\delta$ be equal to $\lambda=\alpha\left[\beta k/\left(\beta+1\right)-1\right]$. As before, upon rescaling time $\lambda t\rightarrow t$, we have $I_{1D}=e^t$, where the subscript 1D stands for the solution of the 1D model.

To find the second constraint on the rates of the 1D model, we calculate $\Pi_n$ -- the extinction probability starting from $n$ infected cells -- in the framework of the 1D model. $\Pi_n$ satisfies the following recursive equation~\cite{Redner2010}
\begin{equation}
\Pi_n=\frac{ n\gamma}{z}\sum_{k=0}^{\infty}D\left(k\right) \Pi _ {n+\left(k-1\right)}+\frac{ n\delta}{z} \Pi _ {n-1},\label{1D extinction probability}
\end{equation}
where $z=n\left(\gamma+\delta\right)$ is the sum of all rates. As before, since the reaction rates are linear in $n$, the solution has the form $\Pi_n=\eta^n$, where $\eta$ is the probability that an infection initiated with a single infected cell will result in extinction. The solution must also satisfy the boundary conditions $\Pi_\infty=0$ and $\Pi_0=1$. Plugging the solution into Eq.~(\ref{1D extinction probability}) we arrive at
\begin{equation}
0=\sum_{k=0}^{\infty}\frac{\gamma}{\gamma+\delta}D\left(k\right)\eta^k-\eta+\frac{\delta}{\gamma+\delta}.\label{rho_1}
\end{equation}
For the case of the $K$-step BSD we find~\cite{Redner2010}
\begin{equation}
0=\eta^{K}-\frac{\gamma+\delta}{\gamma}\eta+\frac{\delta}{\gamma},\label{extinction probability k const}
\end{equation}
while for the geometric BSD the result is $\eta=[\gamma+\delta\left(\langle k\rangle+1\right)]/[\langle k\rangle\left(\gamma+\delta\right)]$.

We now demand that the extinction probability starting with one infected cell in the 1D model be equal to the extinction probability starting with one virion in the 2D model.  The demand comes from the equivalency of the virion dynamics in 2D and the infected cell dynamics in 1D. Indeed, in 2D, a virion can either die or become an infected cell, which then turns with probability 1 to $k$ virions. This is identical to the behavior of an infected cell in 1D. Therefore, since the equation for $\eta$ in the 1D case and the equation for $\rho_V$ in the 2D case are identical, comparing the coefficients in Eqs. (\ref{7}) and (\ref{rho_1}), valid for generic $D(k)$, yields $\delta/\gamma=1/\beta$.
Combined with the demand on the effective growth rate,
we find
\begin{equation}
\gamma=\frac{\beta\alpha}{\beta+1},\quad\delta=\frac{\alpha}{\beta+1}.
\end{equation}
Having found $\gamma$ and $\delta$, we can now analyze the master equation for $P_n\left(t\right)$ in the 1D case:
\begin{equation}
\dot{P_n}(t)=\frac{\alpha}{\lambda\left(\beta\!+\!1\right)}\left[\beta\left(\sum_{k=0}^{\infty}D\left(k\right)\left[n\!-\!k\!+1\right]P_{n-k+1}(t)-n P_n(t)\right)+(n\!+\!1) P_{n+1}(t)- n P_n(t)\right]\!.
\end{equation}
As before, multiplying this equation by $x^n$, and summing over all $n$'s, we arrive at the following evolution equation for the PGF:
\begin{equation}\label{PG1}
\frac{\partial G}{\partial t}=\frac{\alpha}{\lambda\left(\beta+1\right)}\left[ \beta\sum_{k=0}^{\infty}D\left(k\right)x^k-\beta x+1- x\right]\frac{\partial G}{\partial x}.
\end{equation}

Similarly as in the 2D case, by employing the method of characteristics this equation is solvable only for the SSR case, namely $D\left(k\right)=\delta_{k, 2}$. In this case, taking the initial condition $G(x,0)=x$ and boundary condition $G(1,t)=1$, and defining the survival probability starting from one infected cell in the 1D model as $Q_{1D}=1-\eta=1-1/\beta$, we have
\begin{equation}
G\left(x,t\right)=\frac{1-Q_{1D}-e^t (x-1)(Q_{1D}-1)- x}{1-Q_{1D}+e^t (x-1)-x}.\label{GF1D}
\end{equation}
This equation coincides with Eq.~(\ref{13}). Therefore the DIT in 1D coincides with Eq.~(\ref{Time Distribution 2D}) up to the value of the survival probability, which differs between the 2D and 1D models. This discrepancy, however, can be remedied by noticing that the threshold for infection, $N_0$, also varies between the 2D and 1D models. Indeed, in the 2D model $I_{2D}(t)$ grows on average as $\beta/\left(\beta+1\right)e^{t}$ while in the 1D model, $I_{1D}(t)$ grows as $e^{t}$. Therefore, the threshold in the 1D model has to be multiplied by $(\beta+1)/\beta$, in order for the DITs to fully coincide.

\begin{figure}[ht]
	\centering
	\includegraphics[width=.65\linewidth]{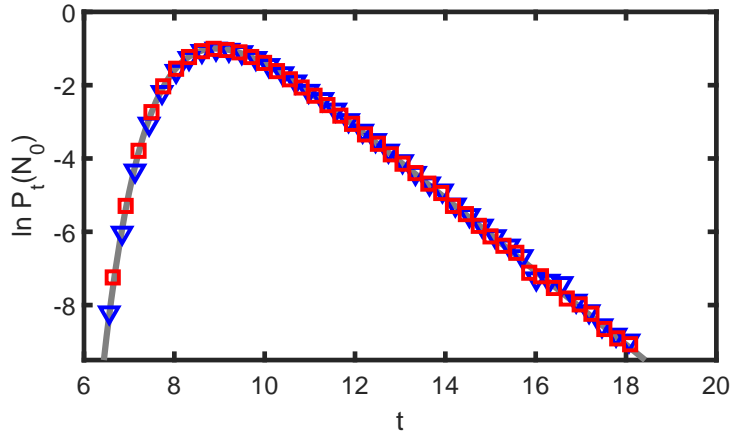}
	\caption{Shown is the distribution of infection times to reach $N_0$ infected cells as a function of the rescaled time, for the SSR. Analytical results (solid line) are compared with results of numerical simulations of the 1D (squares) and 2D (triangles) models, for $\beta=2$, $\alpha=0.01$, and $N_0=10^4$. The fact that the curves are almost indistinguishable indicates that the effective 1D model is an excellent approximation of the 2D model for $\alpha\ll 1$.}
	\label{fig:5}
\end{figure}

We have simulated the 1D model using the Gillespie algorithm, and compared the analytical and numerical distributions. As can be seen in Fig. \ref{fig:5}, both the 1D and 2D numerical distributions excellently agree with the analytical result [Eq.~(\ref{Time Distribution 2D})]. Here and in all other figures showing comparisons between the 1D and 2D models, given a threshold of $N_0$ in the 2D model, we have taken $(\beta+1)N_0/\beta$ as the threshold for the 1D model.

\subsection{Bifurcation limit of the 1D model}\label{BLimit}
Equation~(\ref{PG1}) is exactly solvable only for the SSR case. Yet, close to the bifurcation limit, where $R$ approaches 1, the equation is also approximately solvable, for any $K$. To find the solution, we notice that the extinction probability $\eta$ approaches $1$ for $R-1\ll 1$. Thus, substituting $\eta=1-\delta\eta$ in
Eq.~(\ref{extinction probability k const}), where $\delta\eta\ll 1$, and solving the resulting equation in the leading order in $\delta\eta$, yields the survival probability $Q_{1D}=\delta\eta=2(R-1)/(K-1)$.

\begin{figure}[ht]
	\centering
	\includegraphics[width=.9\linewidth]{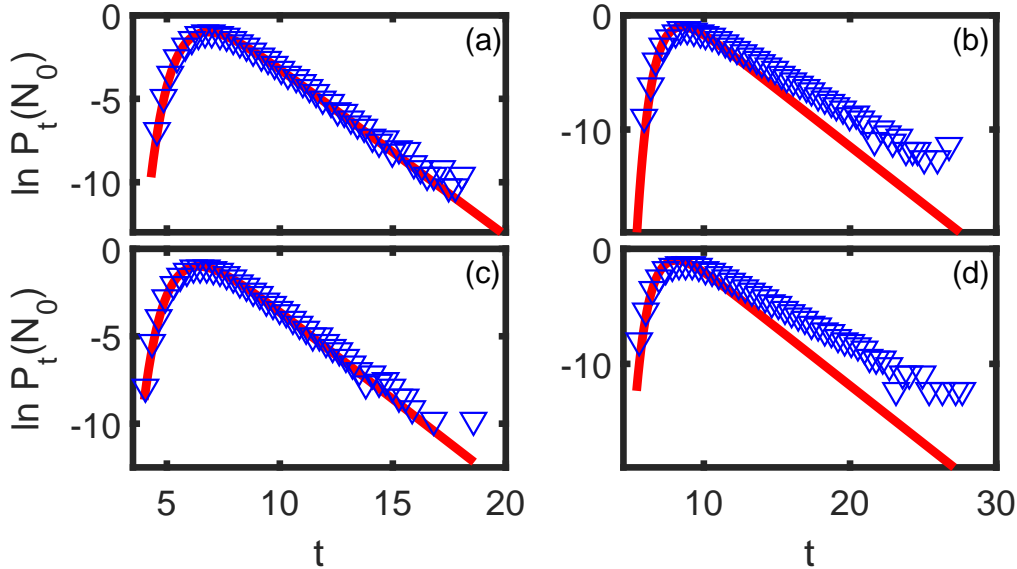}
	\caption{Shown are distributions of infection times to reach $N_0$ infected cells as a function of the rescaled time, for the $K$-step BSD. The analytical result close to bifurcation (solid line) is compared with numerical simulations (triangles) for different $K$ and $R$ values: (a) $K=3$ and $R=1.1$; (b) $K=3$ and $R=2$; (c) $K=4$ and $R=1.1$; (d) $K=4$ and $R=2$. One can see that the analytical and numerical results agree well as long as $K\left(R-1\right)\ll 1$, see text. Here $\alpha=0.01$, $N_0=10^4$, and $\beta$ is determined by the $K$ and $R$ values. }
	\label{fig:6}
\end{figure}

To solve Eq.~(\ref{PG1}) close to the bifurcation limit, we substitute $D(k)=\delta_{k,K}$, and expand the coefficients of $\partial G/\partial x$ around $x=1$ up to second order in $\left|1-x\right|\ll1$, which yields
\begin{equation}\label{Gbif}
\frac{\partial G}{\partial t}\simeq \left[\left(x-1\right)+\frac{1}{Q_{1D}}\left(x-1\right)^2\right]\frac{\partial G}{\partial x}.
\end{equation}
It can be shown that this approximation holds as long as $K(R-1)\ll 1$. Note, that it is justified to expand Eq.~(\ref{PG1}) in the vicinity of $x=1$, as close to the bifurcation limit, the probability distribution at long times is determined by a narrow region of $G$ in the vicinity of $x=1$~\cite{assaf2006spectral,assaf2007}. In fact, what we are doing here is approximating the $K$-step process by a single-step process $I\rightarrow2I$ [for which Eq.~(\ref{Gbif}) is exact], which turns out to be a good approximation when $R\rightarrow 1$. Since Eq.~(\ref{Gbif}) coincides with Eq.~(\ref{PG1}) upon substituting $D(k)=\delta_{k,2}$ and using the definition of $Q_{1D}$ close to the bifurcation limit, we find that in this limit, the DIT coincides with Eq.~(\ref{Time Distribution 2D}) up to the value of the survival probability $Q_{1D}$, given above Eq.~(\ref{Gbif}).
In Fig.~\ref{fig:6} we compare the analytical result close to bifurcation with numerical simulations for $K=3$ and $K=4$. For each $K$, we have taken two $R$ values: $R=1.1$ and $R=2$. As shown in the figure, the DIT close to bifurcation is accurately given by the SSR result, as long as $K(R-1)\ll 1$.

\section{Numerical Analysis}\label{Numerical Analysis}
\subsection{Monte-Carlo simulation results}\label{Comparison 1D 2D Models}
\begin{figure}[ht]
	\centering
	\includegraphics[width=.9\linewidth]{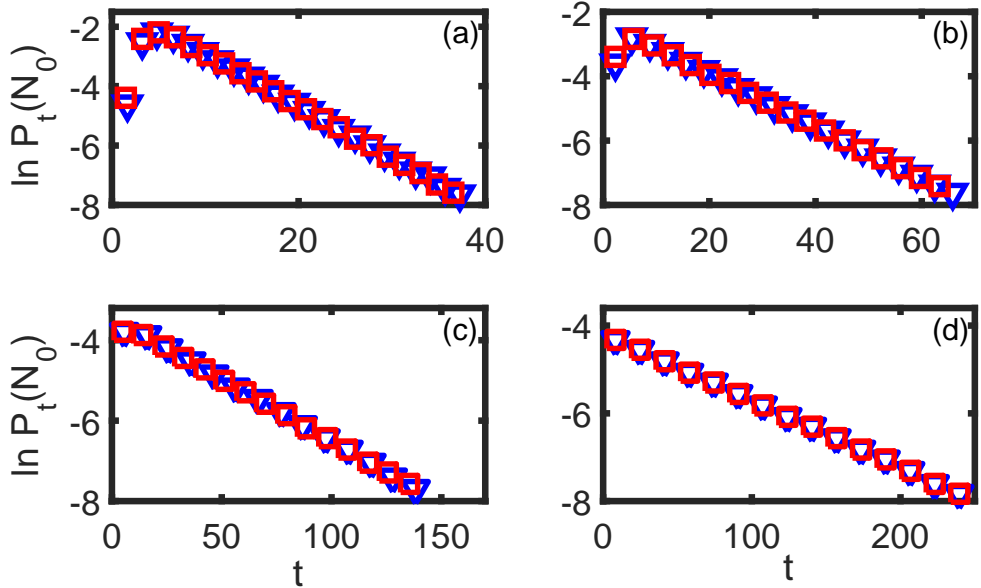}
	\caption{Shown are distributions of infection times to reach $N_0$ infected cells as a function of the rescaled time, for the $K$-step BSD. Here $\beta=2$, $N_0=100$, $\alpha=0.01$ and (a) $K=10$, (b) $K=20$, (c) $K=50$, (d) $K=100$. The numerical results of the 2D model (triangles) agree well with those of the effective 1D model (squares) for all $K$ values. }
	\label{fig:7}
\end{figure}

\begin{figure}[ht]
	\centering
	\includegraphics[width=.7\linewidth]{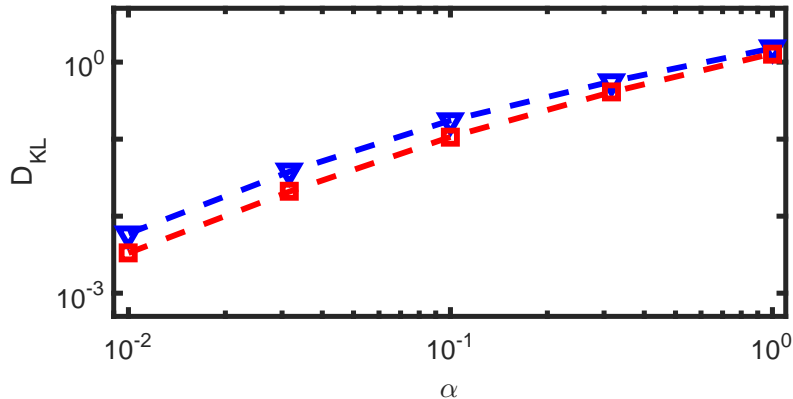}
	\caption{Shown is the Kullback-Leibler divergence (see text) between the numerically-calculated distributions of infection times, of the 1D and 2D models, for the $K$-step (triangles) and  geometric (squares) BSDs, versus $\alpha$. One can see that, as long as $\alpha\ll 1$, the 1D model remains a good approximation of the 2D model. Here $\beta=0.3$ and $K=\langle k\rangle=50$. }
	\label{fig:8}
\end{figure}

After having compared the 1D and 2D models, both analytically and numerically, for the SSR case and close to bifurcation, we now compare the models numerically for generic BSDs. An example can be seen in Fig. \ref{fig:7}, where we have compared the models for the case of $K$-step BSD with four different $K$'s, demonstrating an excellent agreement between the models.
A method that is commonly used to measure the similarity between two distributions is called the  Kullback-Leibler (KL) divergence \cite{kullback1951information}, defined by
\begin{equation}
D_{KL}\left(p\|q\right)=\sum_{i=1}^{N}p\left(x_i\right)\log{\frac{p\left(x_i\right)}{q\left(x_i\right)}}.
\end{equation}
Here $p$ is a probability distribution and $q$ is an approximated probability distribution to $p$. In general, as $D_{KL}$ approaches zero, the probability distribution of $q$ approaches $p$, whereas when $D_{KL}=\mathcal{O}\left(1\right)$, $q$ is a poor approximation of $p$.  The results of the  KL divergence  between the 1D and 2D models are shown in Fig. \ref{fig:8}. We have seen that the time-scale separation leading to the effective 1D model requires that $\alpha\ll 1$. As a result, we expect the 1D approximation to break down as $\alpha$ becomes ${\cal O}(1)$. In Fig. \ref{fig:8}  the KL divergence is shown as a function of $\alpha$ for both the $K$-step and the geometric BSDs. It is evident that as $\alpha$ increases the approximation deteriorates, whereas for $\alpha=\mathcal{O}(1)$, $D_{KL}$ becomes $\mathcal{O}(1)$, which indicates that the 1D approximation breaks down in this regime.

\subsection{Inferring the statistics of infection times}\label{Statistics Infection}
Previously we have analytically computed the DIT for the SSR case. We now present a semi-empirical argument that allows finding the DIT for a generic choice of BSD. To do so, let us observe how the mean and variance of the DIT behave as a function of the BSD. In the SSR case, computing the mean and standard deviation (STD) is straightforward. The mean time to infection is given by
$\langle t\rangle=\int_0^{\infty}tP_t\left(N_0\right)dt\simeq \ln (Q_{2D}N_0)$, where $P_t(N_0)$ was given from Eq.~(\ref{Time Distribution 2D}), and the approximation holds when $\langle t\rangle\gg1$. Furthermore, the variance of the DIT in this case satisfies, $\sigma^2=\int_0^{\infty}t^2P_t(N_0)\mathrm{d}t - \langle t\rangle^2\simeq\pi^2/6$.

For $K>2$, $P_t(N_0)$ is unknown analytically, and thus the mean and variance cannot be found in this way. However, we have empirically observed that for large $K$, the right tail of the DIT decays exponentially with a slope that equals $1/(R-1)$. At the same time, these distributions have a bulk region, around the maximum, obtained approximately at $t=\ln N_0$. Thus, when $K$ is large, both the mean and variance of the DIT are governed by the exponentially decaying right tail, namely, they are solely determined by $R$, and are independent on other parameters. Using the slope of the right tail which we found empirically, the mean becomes $\langle t_{K-step}\rangle\simeq\ln N_0+R-1$, while the STD satisfies $\sigma_{K-step}\simeq R-1$. When the BSD is geometrically distributed, we have similarly found that $\langle t_{geometric}\rangle\simeq\ln N_0+R$ and $\sigma_{geometric}\simeq R$.
\begin{figure}[ht]
	\centering
	\includegraphics[width=.6\linewidth]{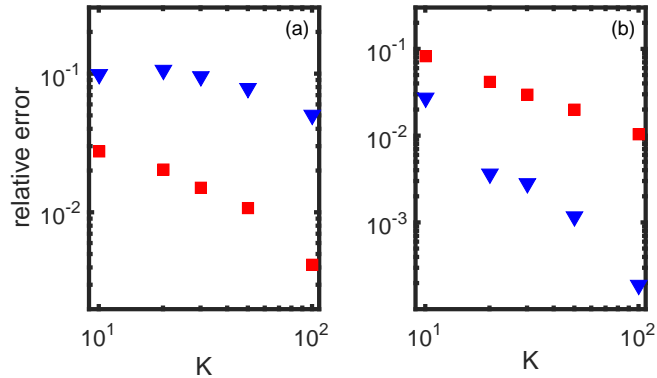}
	\caption{(a) The relative error (see text) between the theoretical expression for the mean time to infection, $\langle t\rangle_{theo}$, and the numerical result, $\langle t\rangle_{num}$, as a function of $K$, for $\beta=2$. (b) The relative error between the theoretical expression for the STD of the distribution of infection times, $\sigma_{theo}$, and the numerical result, $\sigma_{num}$, as a function of $K$, for $\beta=2$. In both panels the triangles represent the $K$-step BSD and the squares represent the geometric BSD, with $\langle k\rangle=K$.  }
	\label{fig:9}
\end{figure}

Figure \ref{fig:9} presents the relative error between the estimated expressions for the mean  and STD, and the numerical results. We denote by $\langle t\rangle_{theo}$ and $\sigma_{theo}$ the theoretical mean and STD respectively, while $\langle t\rangle_{num}$ and $\sigma_{num}$ respectively denote the numerical mean and STD of the DIT. The relative error was calculated by computing $\left|\langle t\rangle_{theo}-\langle t\rangle_{num}\right|/\langle t\rangle_{num}$, and $\left|\sigma_{theo}-\sigma_{num}\right|/\sigma_{num}$. One can see a good agreement between the semi-empirical and numerical results, which improves as
$K$ (and correspondingly $R$) is increased, where the tail of the DIT becomes closer to an exponential distribution.

\begin{figure}[ht]
	\centering
	\includegraphics[width=.65\linewidth]{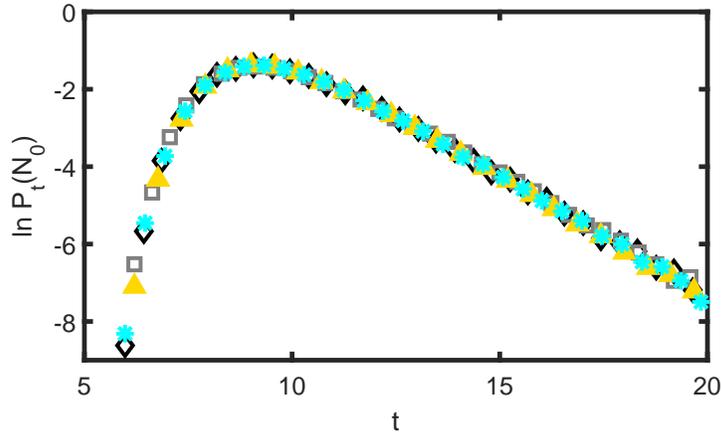}
	\caption{Shown are distributions of infection times to reach $N_0$ infected cells, as a function of the rescaled time, for the $K$-step BSD. Here $\alpha=0.01$, and the different symbols represent $K=10$ $(\triangle)$, $K=15$ $(*)$, $K=20$ $(\diamond)$ and $K=50$ $(\square)$, while we chose values of $\beta$ such that $R=1.9$ and $\lambda=0.009$ for all the curves. One can see that despite changing $K$ by a factor of $5$, the curves are almost indistinguishable, indicating that the distribution is solely governed by $R$, see text.}
	\label{fig:10}
\end{figure}

To verify our empirical result that for large $K$ the DITs depend only on $R$, we plotted in Fig.~\ref{fig:10} several DITs by varying both $K$ and $\beta$ such that $R$ remained constant. This figure clearly demonstrates that the DITs are governed by a single parameter when $K$ is large.

\begin{figure}[ht]
	\centering
	\includegraphics[width=.6\linewidth]{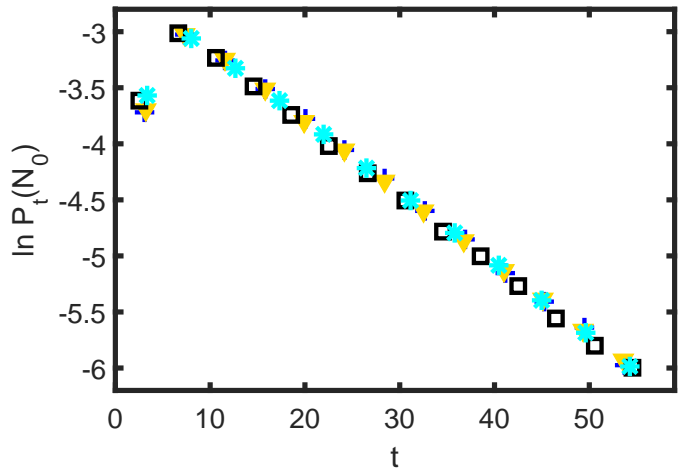}
	\caption{Shown are distributions of infection times to reach $N_0$ infected cells, as a function of the rescaled time. Here the result for the $K$-step BSD $(*)$ is compared with results of bimodal BSDs with a width of $5$ $(+)$, $10$ $(\square)$, and $50$ $(\bigtriangledown)$, for $R=16.66$, $K=50$, $\beta=0.5$, $\alpha=0.01$, and $N_0=10^3$. The figure indicates that at large values of $K$, the distribution of infection times is insensitive to the higher moments of the BSD (such as its width). }
	\label{fig:11}
\end{figure}

Having shown that for the $K$-step BSD the DIT is solely governed by $R$ (as long as $K\gg1$), we wanted to check whether the \textit{width} of the BSD affects the DIT. To do so, we took a bi-modal BSD with $D(k)=1/2\delta_{k,K-\Delta}+1/2\delta_{k,K+\Delta}$, where $\Delta$ is the width of the BSD. For each BSD used, we kept the mean $K$ constant while changing its width (by changing $\Delta$). The results were then compared with the DIT in the case of the $K$-step BSD, and we found that the BSDs were independent on $\Delta$, see Fig. \ref{fig:11}. These results indicate that when $K$ is large, the DIT only depends on the BSD's mean (through $R$). That is, the DIT is independent on the higher moments of the BSD  such as its width. This result also holds for the geometric BSD as discussed above.

\subsection{Implications on Realistic Populations}\label{Implications Realistic Population}
We now show which possible implications our semi-empirical results have on the estimation of the DIT, using realistic biological parameters, for both the HIV and HCV.

\begin{figure}[ht]
	\centering
	\includegraphics[width=.9\linewidth]{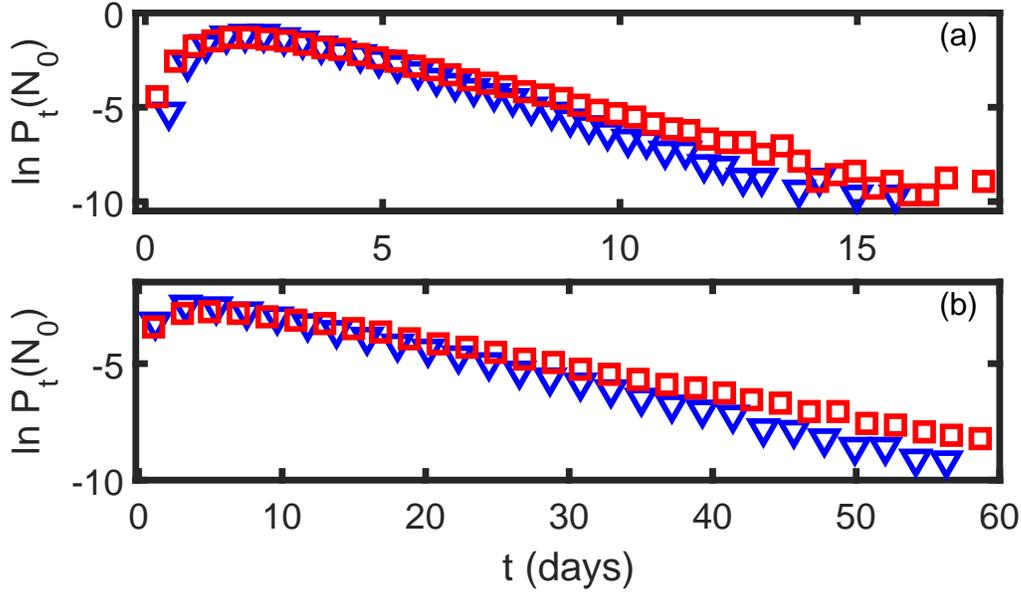}
	\caption{Shown are distributions of infection times to reach $N_0$ infected cells, versus the number of infection days. (a) HIV with $\beta=0.15$, $\alpha=0.05$, $\langle k\rangle=K=20$ and $N_0=32$, such that $N_0^{1D}=245$. (b) HCV with $\beta=0.125$, $\alpha=0.0175$, $\langle k\rangle=K=50 $ and $N_0^{1D}=110$. In both panels the triangles represent the $K$-step BSD and the squares represent the geometric BSD. Note, that in both cases the distribution displays a fat right tail, see text.}
	\label{fig:12}
\end{figure}

As a first example, let us take a set of realistic rates that has been studied in \cite{pearson2011stochastic} for the HIV, where $\beta=0.15$, $\alpha=0.05$ and $K=20$, for which $R\simeq 2.6$.
Pearson et al. \cite{pearson2011stochastic} have shown that for this set of parameters taking $N_0=32$ guarantees persistent infection, with a vanishingly low extinction probability. We have numerically computed the DIT for this set of parameters, for the cases of $K$-step and geometric BSDs, with $\langle k \rangle=K$, see Fig.~\ref{fig:12}~\cite{footnote5}.
With these parameters the mean time to infection is $\langle t\rangle\simeq 3$ days for both BSDs, while the  STDs are $\sigma_{K-step}\simeq 1.5$ and $\sigma_{geometric}\simeq 1.9$ days. Notably, while 3 days is the mean infection time, in the case of the $K$-step BSD, approximately 5$\%$ of the population will only be infected after 6 days, and 0.2$\%$ after 10 days. For the geometric BSD, approximately 8$\%$ of the population will be infected after 6 days, and 0.7$\%$ after 10 days. These numbers are much higher than those obtained for a Gaussian DIT with the same mean and STD.

As a second example, we took HCV (hepatitis C virus), with realistic parameters of $\alpha=0.0175$, $K=50$ and $\beta=0.125$~\cite{neumann1998hepatitis}. %,ramratnam1999rapid,dahari2005mathematical,chang2003dynamics,dahari2009mathematical}.
In Fig.~\ref{fig:12} we also plotted the DIT for the HCV and found that the mean time for infection satisfies $\langle t_{K-step}\rangle\simeq 9.5$ and $\langle t_{geometric}\rangle\simeq 11.9$ days, while the STDs are $\sigma_{K-step}\simeq 7.6$ and $\sigma_{geometric}\simeq 9.5$ days.
Importantly, our results indicate that there is a non-negligible fraction of the population whose infection time is significantly longer than the typical infection time, by many standard deviations, and thus, it is imperative to know the entire DIT. That is, the fact that the DIT is highly-skewed, for both the HIV and HCV, clearly demonstrates that knowing the bulk of the DIT, described by the mean infection time and its standard deviation, is insufficient for assessing the infection times of a significant portion of the population.

\section{Conclusions and Discussion}\label{Conclusion Discussion}
We have investigated the dynamics of the early infection stage of viral diseases such as HIV. The model we have considered included the dynamics of both the virions and infected cells. By employing the probability generating function formalism  we were able to analytically find the distribution of infection times (DIT) for a particular choice of virion burst size distribution (BSD), while for other choices of BSD we have computed the DIT numerically and semi-empirically. Furthermore, by exploiting the time-scale separation between the dynamics of the virions and infected cells, we were able to reduce the 2D model into an effective 1D model for the infected cells only. Using the measure of Kullback-Leibler divergence, we have shown that the results for the DIT in the 1D model coincide with those of the 2D model, for any choice of BSD, in the limit of fast virion dynamics. We have also considered the bifurcation limit where the infected cell population grows slowly, and found the DIT for any arbitrary BSD.

Our results for the DIT indicate that for a realistic choice of parameters, the right tail of the DIT is exponential and thus, it is skewed towards the right. Therefore, the bulk of the DIT, described by the mean infection time and its standard deviation, is a poor measure for assessing whether an individual has been infected or not. Importantly, this may have implications on when HIV (or any other viral infection that behaves similarly) tests should be performed.

Notably, while our model is generic and holds for a wide variety of viruses, it is valid only for the early stages of infection, as it neglects immune system responses and medical treatment, such as vaccinations or medications. It also neglects other types of interactions between viruses and healthy T cells via apoptosis (self-destruction)~\cite{selliah2003t}, and cell to cell transmission of the virus, which may be significantly more efficient than infection by virions~\cite{dimitrov1993quantitation,galloway2015cell,murooka2012hiv}. It would be interesting to study the implications of these various factors on the statistics of infection times.

\section{Acknowledgements}
This work was supported through the Israel Science Foundation Grant No. 300/14 and the United States-Israel Binational Science Foundation grant No. 2016-655.

%%%%%%%%%%%%%%%%   bib   %%%%%%%%%%%%%%%%%%%%%%%%%%%%%%%%


\begin{thebibliography}{9}
\bibitem{footnote2} Although the HCV replicates mainly in hepatocytes, cells of the liver, and the influenza virus replicates in the epithelial cells throughout the respiratory tree, their dynamic is similar to that of the HIV.
\bibitem{footnote3} Note that while non-linear saturation terms are negligible for small populations, they become important for large populations, which can already be recognized by the immune system~\cite{mcmichael2010immune}. %Nevertheless, extinction of the virion and infected cell populations can occur even without including such non-linear terms.
\bibitem{footnote5} Here, one can see that the right tail of the geometric BSD is fatter than that of the K-step BSD. A possible explanation is that when a geometric BSD is used, the system can sample bursts of size $k$ that are much larger than $\langle k\rangle$, and thus, the tail becomes fatter.
\end{thebibliography}
\end{document}